\documentclass[11pt,onecolumn,draftclsnofoot]{IEEEtran}
\usepackage[noadjust]{cite}

\ifCLASSINFOpdf
   \usepackage[pdftex]{graphicx}
   \DeclareGraphicsExtensions{.pdf,.jpeg,.png}
\else
   \usepackage[dvips]{graphicx}
   \DeclareGraphicsExtensions{.eps}
\fi
\usepackage[cmex10]{amsmath}
\usepackage{amssymb,bm,cases,subfigure}
\usepackage{color,pstricks}
\usepackage{sgame}
\usepackage{egame}
\newcommand{\Tr}{\mathrm{Tr}}

\begin{document}
%
\title{Jamming Games in the MIMO Wiretap Channel With an Active Eavesdropper}

\author{Amitav~Mukherjee and A.~Lee~Swindlehurst
\thanks{Copyright (c) 2012 IEEE. Personal use of this material is permitted. However, permission to use this material for any other purposes must be obtained from the IEEE by sending a request to pubs-permissions@ieee.org.}
\thanks{The authors are with the Dept.~of EECS,
University of California, Irvine, CA 92697, and acknowledge support from the U.S. Army Research Office MURI grant W911NF-07-1-0318. {\tt (e-mail: \{amukherj; swindle\}@uci.edu)} }
}


\maketitle

\begin{abstract}
This paper investigates reliable and covert transmission strategies in
a multiple-input multiple-output (MIMO) wiretap channel with a
transmitter, receiver and an adversarial wiretapper, each equipped
with multiple antennas. In a departure from existing work, the
wiretapper possesses a novel capability to act either as a passive
eavesdropper or as an active jammer, under a half-duplex
constraint. The transmitter therefore faces a choice between
allocating all of its power for data, or broadcasting artificial interference
along with the information signal in an attempt to jam the
eavesdropper (assuming its instantaneous channel state is unknown). To
examine the resulting trade-offs for the legitimate transmitter and
the adversary, we model their interactions as a two-person zero-sum
game with the ergodic MIMO secrecy rate as the payoff function.
We first examine conditions for
the existence of pure-strategy Nash equilibria (NE) and the structure of mixed-strategy NE for the strategic form
of the game. We then derive equilibrium strategies for the extensive form of the game where
players move sequentially under scenarios of perfect and imperfect information. Finally, numerical simulations are
presented to examine the equilibrium outcomes of the various scenarios considered.
\end{abstract}

\begin{IEEEkeywords}
Physical layer security, MIMO wiretap channel, game theory, jamming, secrecy rate, Nash Equilibria.
\end{IEEEkeywords}

\section{INTRODUCTION}
The two fundamental characteristics of the wireless medium, namely \emph{broadcast} and \emph{superposition}, present different challenges in ensuring secure and reliable communications in the presence of adversaries. The broadcast nature of wireless communications makes it difficult to shield transmitted signals from unintended recipients, while superposition can lead to the overlapping of multiple signals at the receiver. As a result, adversarial users are commonly modeled either as (1) a passive \emph{eavesdropper} that tries to listen in on an ongoing transmission without being detected, or (2) a malicious transmitter (\emph{jammer}) that tries to degrade the signal quality at the intended receiver. Two distinct lines of research have developed to analyze networks compromised by either type of adversary, as summarized below.

A network consisting of a transmitter-receiver pair and a passive eavesdropper is commonly referred to as the {\em wiretap} channel. The information-theoretic aspects of this scenario have been explored in some detail \cite{Wyner75,Csiszar78,LeungH78}. In particular, this work led
to the development of the notion of {\em secrecy capacity}, which
quantifies the maximal rate at which a transmitter can reliably send a secret
message to the receiver, without the eavesdropper being able to decode
it. Ultimately, it was shown that a non-zero secrecy capacity can only
be obtained if the eavesdropper's channel is of lower quality than
that of the intended recipient. The secrecy capacity metric for the multiple-input multiple-output (MIMO) wiretap channel, where
all nodes may possess multiple antennas, has been studied in
\cite{OggierH08}-\cite{Wornell09}, for example. There are two primary categories of secure transmission strategies for the MIMO wiretap channel, depending on whether the instantaneous channel realization of the eavesdropper is known or unknown at the transmitter. In this work we assume that this information is not available, and thus the transmitter incorporates an ``artificial interference'' signal \cite{GoelN08}-\cite{MukherjeeTSP} along with the secret message in an attempt to degrade the eavesdropper's channel, as elaborated on in Section~\ref{sec:model}.

The impact of malicious jammers on the quality of a communication link is another problem of long-standing interest, especially in mission-critical and military networks. A common approach is to model the transmitter and the jammer as players in a game-theoretic formulation with the mutual information as the payoff function, and to identify the optimal transmit strategies for both parties \cite{Stark88}-\cite{Diggavi01}. Recent work has extended this technique to compute the optimal spatial power allocation for MIMO and relay channels with various levels of channel state information (CSI) available to the transmitters \cite{Basar04}-\cite{Giannakis08}.

In this paper, we consider a MIMO communication link in the presence
of a more sophisticated and novel adversary, one with the dual capability of
either passively eavesdropping or actively jamming any ongoing
transmission, with the objective of causing maximum disruption to the
ability of the legitimate transmitter to share a secret message with
its receiver. The legitimate transmitter now faces the dilemma of
establishing a reliable communication link to the receiver that is
robust to potential jamming, while also ensuring confidentiality from
interception. Since it is not clear \emph{a priori} what strategies
should be adopted by the transmitter or adversary per channel use, a
game-theoretic formulation of the problem is a natural solution due to
the mutually opposite interests of the agents. Unlike the jamming
scenarios mentioned above that do not consider link security, the game
payoff function in our application is chosen to be the ergodic
\emph{MIMO secrecy rate} between the legitimate transmitter-receiver
pair.  Related concurrent work on the active eavesdropper scenario
\cite{Amariucai09,Erkip09} has focused on single-antenna nodes without
the use of artificial interference, possibly operating together with
additional `helping' relays.  The
single-antenna assumption leads to a much more restrictive set of user
strategies than the MIMO scenario we consider.

The contributions of the paper are as follows: (1) we show how to
formulate the MIMO wiretap channel with a jamming-capable eavesdropper
as a two-player zero-sum game, (2) we characterize the conditions
under which the strategic version of the game has a pure-strategy Nash equilibrium,
(3) we derive the optimal mixed strategy profile for the players when
the pure-strategy Nash equilibrium does not exist, and (4) we study the extensive or
Stackelberg version of the game where one of the players moves first
and the other responds, and we also characterize the various
equilibrium outcomes for this case under perfect and imperfect information.  These contributions appear in the paper as follows.
The assumed system model and CSI assumptions are presented in the next section.
The strategic formulation of the wiretap game
is described in Section~\ref{sec:strats}, where the two-player
zero-sum payoff table is developed, the conditions for existence of pure-strategy
Nash equilibria are derived, and the optimal mixed strategy
formulation is discussed.  The extensive version of the wiretap game
with perfect and imperfect information where the players move
sequentially is detailed in Section~\ref{sec:extensive}.  Outcomes for
the various game formulations are studied via simulation in
Section~\ref{sec:sim}, and conclusions are presented in
Section~\ref{sec:concl}.

\emph{Notation}:
We will use $\mathcal{CN}(0,1)$ to denote a circular complex
Gaussian distribution with zero-mean and unit variance. We also use
$\mathcal{E}\{\cdot\}$ to denote expectation, $I(\cdot ;\cdot)$ for mutual
information, $(\cdot)^T$ for the transpose, $(\cdot)^H$ for the Hermitian
transpose, $(\cdot)^{-1}$ for the matrix inverse, $\Tr(\cdot)$ for the
trace operator, $\left| \cdot \right|$ to denote the matrix determinant, $\lambda_i(\mathbf{A})$ is the $i^{th}$ ordered eigenvalue of $\mathbf{A}$, and
$\mathbf{I}$ represents an identity matrix of appropriate dimension.

\section{SYSTEM MODEL}\label{sec:model}

\begin{figure}[htp]
\centering
\includegraphics[width=\linewidth]{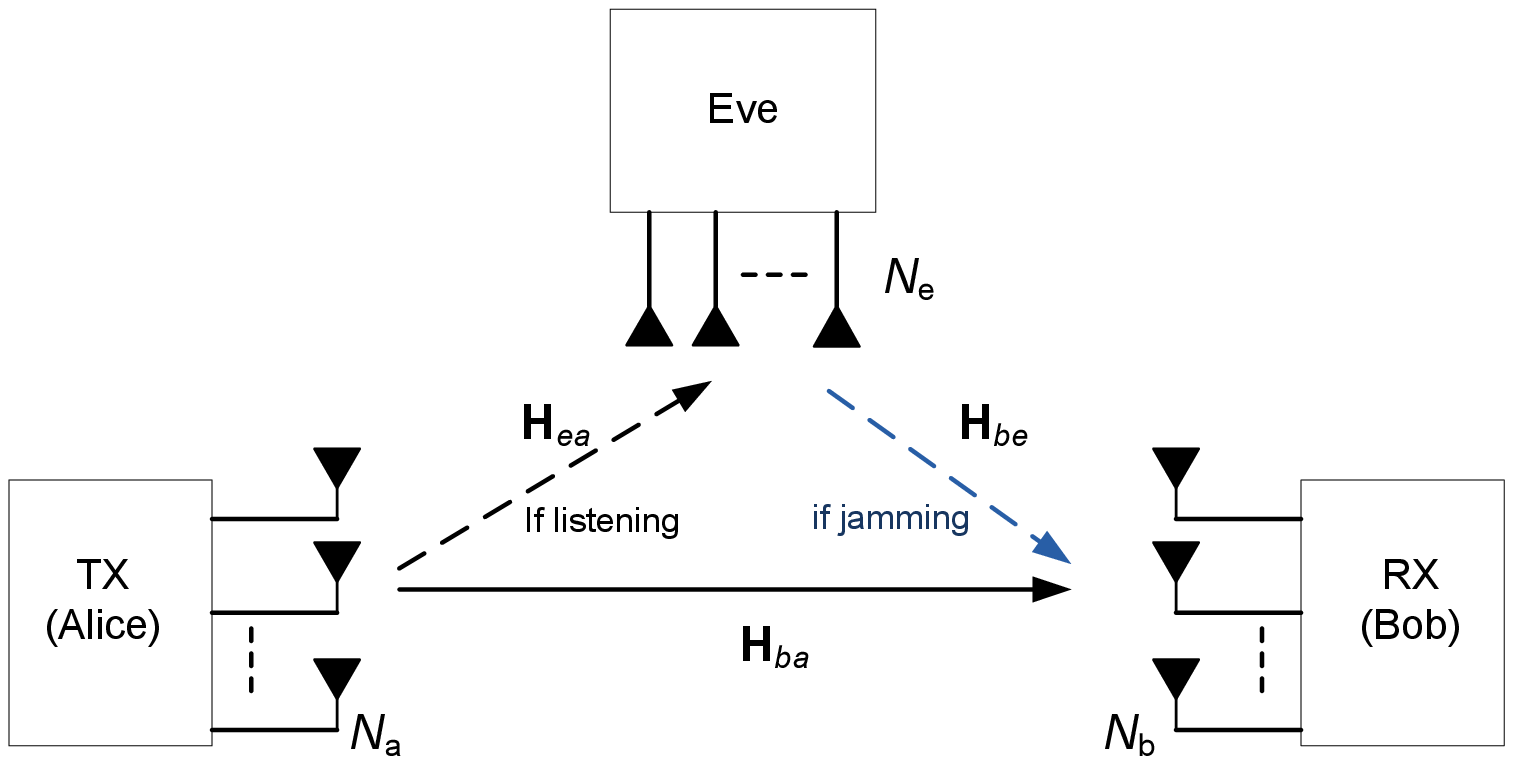}
\caption{MIMO wiretap channel with dual-mode active eavesdropper.}
\label{fig_MIMOwiretap}
\end{figure}

We study the MIMO wiretap problem in which three multiple-antenna
nodes are present: an $N_a$-antenna transmitter (Alice), an
$N_b$-antenna receiver (Bob), and a malicious user (Eve) with $N_e$
antennas, as shown in Fig.~\ref{fig_MIMOwiretap}. We assume that Alice does not have knowledge of the
instantaneous CSI of the eavesdropper, only the statistical
distribution of its channel, which is assumed to be zero-mean with a
scaled-identity covariance. The lack of instantaneous eavesdropper CSI
at Alice precludes the joint diagonalization of the main and
eavesdropper channels \cite{Wornell09}. Instead, as we will show,
Alice has the option of utilizing all her power for transmitting data
to Bob, regardless of channel conditions or potential eavesdroppers,
or optimally splitting her power and simultaneously transmitting the
information vector and an ``artificial interference'' signal that jams
any unintended receivers other than Bob. The artificial interference
scheme does not require knowledge of Eve's instantaneous CSI, which
makes it suitable for deployment against passive eavesdroppers
\cite{Wornell09,GoelN08},\cite{Swindlehurst09}--\cite{Zhou09}. Eve
also has two options for disrupting the secret information rate
between Alice and Bob: she can either eavesdrop on Alice or jam Bob,
under a half-duplex constraint.

\subsection{Signal Model}
 When Eve is in passive eavesdropping mode, the signal received by Bob is
\begin{equation}
\mathbf{y}_b  =  \mathbf{H}_{ba} \mathbf{x}_a + \mathbf{n}_b, \label{eq:yb_Evepass}
\end{equation}
where $\mathbf{x}_a$ is the signal vector transmitted by Alice,
$\mathbf{H}_{ba}$ is the $N_b\times N_a$ channel matrix between Alice
and Bob with i.i.d elements drawn from the complex Gaussian
distribution $\mathcal{CN}(0,1)$, and $\mathbf{n}_b$ is additive
complex Gaussian noise. When Eve is not jamming, she receives
\begin{equation}
{\mathbf{y}}_e = \sqrt{g_1}{{\mathbf{H}}_{ea}}{{\mathbf{x}}_a} + {{\mathbf{n}}_e},
\end{equation}
where $\mathbf{H}_{ea}$ is the $N_e\times N_a$ channel matrix between
Alice and Eve with i.i.d elements drawn from the complex Gaussian
distribution $\mathcal{CN}(0,1)$, and $\mathbf{n}_e$ is additive
complex Gaussian noise. The background noise at all receivers is
assumed to be spatially white and zero-mean complex Gaussian:
$\mathcal{E}\{\mathbf{n}_k\mathbf{n}_k^H\} = \sigma_k^2 \mathbf{I}$,
where $k=b,e$ indicates Bob or Eve, respectively. The receive and
transmit channels of the eavesdropper have gain factors $\sqrt{g_1}$
and $\sqrt{g_2},$ respectively. These scale factors may be interpreted
as an indicator of the relative distances between Eve and the other
nodes.

On the other hand, when Eve decides to jam the legitimate channel,
Bob receives
\begin{equation}
\mathbf{y}_b  =  \mathbf{H}_{ba} \mathbf{x}_a + \sqrt{g_2}\mathbf{H}_{be} \mathbf{x}_e+ \mathbf{n}_b, \label{eq:yb_Evejam}
\end{equation}
where $\mathbf{x}_e$ is the Gaussian jamming signal from Eve and
$\mathbf{H}_{be}$ is the $N_b\times N_e$ channel matrix between Eve
and Bob with i.i.d elements distributed as $\mathcal{CN}(0,1)$.  Due to
the half-duplex constraint, Eve receives no signal when she is jamming
($\mathbf{y}_e=0$).

Alice's transmit power is assumed to be bounded by $P_a$:
\begin{equation*}
\mathcal{E}\{\mathbf{x}_a\mathbf{x}_a^H\} = \mathbf{Q}_a \qquad
\mbox{\rm Tr}(\mathbf{Q}_a) \le P_a \; ,
\end{equation*}
and similarly Eve has a maximum power constraint of $P_e$ when
in jamming mode.  To cause maximum disruption to Alice and Bob's
link, it is clear that Eve will transmit with her full available power
$P_e$ when jamming. In the most general scenario where Alice jams Eve by
transmitting artificial interference, we have
\begin{equation}
{\mathbf{x}}_a  = {\mathbf{Tz}} + {\mathbf{T'z'}},
\end{equation}
where $\mathbf{T},\mathbf{T'}$ are the $N_a \times d$, $N_a \times (N_a-d)$ precoding matrices for the $d\times 1$ information vector $\mathbf{z}$ and uncorrelated $(N_a-d) \times 1$ jamming signal $\mathbf{z'},$ respectively.  To ensure that the artificial interference does not interfere with the information signal, a common approach taken in the literature \cite{Wornell09,GoelN08},\cite{Swindlehurst09}--\cite{MILCOM10} is to make these signals orthogonal when received by Bob. If Alice knows $\mathbf{H}_{ba}$, this goal can be achieved by choosing $\mathbf{T}$ and $\mathbf{T'}$ as disjoint sets of the right singular vectors of $\mathbf{H}_{ba}$. Note that if the users have only a single antenna, the effect of the  artificial interference cannot be eliminated at Bob, and it will degrade the SNR of both Bob and Eve.  This makes it unlikely that Alice will employ a non-zero artificial interference signal when she has only a single transmit antenna, which significantly restricts Alice's transmission strategy. The matrix $\mathbf{Q}_a$ may be expressed as
\begin{equation}
\mathbf{Q}_a = \mathbf{T}\mathbf{Q}_z\mathbf{T}^H + \mathbf{T}'\mathbf{Q}'_z\mathbf{T}'^H , \label{QT}
\end{equation}
where $\mathbf{Q}_z, \mathbf{Q}'_z$ are the covariance matrices
associated with $\mathbf{z}$ and $\mathbf{z}'$, respectively.  If we
let $\rho$ denote the fraction of the total power available at Alice
that is devoted to the information signal, then $\mbox{\rm
Tr}(\mathbf{T}\mathbf{Q}_z\mathbf{T}^H)= \rho P_a$ and $\mbox{\rm
Tr}(\mathbf{T}'\mathbf{Q}'_z\mathbf{T}'^H) = (1-\rho)P_a$.  Due
to the zero-forcing constraint, it is clear that any power available
to Alice that is not used for the desired signal $\mathbf{x}_a$ will
be used for jamming, so between the signal and artificial
interference, Alice will transmit with full power $P_a$.  The
covariance matrices of the received interference-plus-noise at Bob and
Eve are
\begin{eqnarray}
{{\mathbf{K}}_b} & = & \left\{ {\begin{array}{*{20}{c}}
  {{g_2}{{\mathbf{H}}_{be}}{{\mathbf{Q}}_{be}}{\mathbf{H}}_{be}^H + \sigma _b^2{\mathbf{I}}}&{{\text{if Eve jams}}} \\
  {\sigma _b^2{\mathbf{I}}}&{{\text{if Eve listens}}}
\end{array}} \right.\\
\mathbf{K}_e & = & g_1\mathbf{H}_{ea}\mathbf{T}'\mathbf{Q}'_z \mathbf{T}'^H \mathbf{H}_{ea}^H +\sigma_e^2 \mathbf{I}, \label{eq:Qb_Qe}
\end{eqnarray}
where $\mathbf{Q}_{be}$ is the covariance of the jamming signal transmitted by Eve, $\Tr(\mathbf{Q}_{be}) \le P_e$.

Note that we have assumed that Alice's jamming signal (if any) is orthogonal to the information signal received by Bob, and hence, from the point of view of mutual information, can be ignored in the expression for $\mathbf{K}_b$.
For our purposes, we assume that Alice splits her transmit power between a stochastic encoding codebook and artificial interference for every channel use in \emph{all} scenarios, while Bob employs a deterministic decoding function \cite{Wyner75,Csiszar78}. Firstly, this ensures that the general encoding and decoding architecture of the Alice-Bob link remains fixed irrespective of Eve's actions. Secondly, for a point-to-point channel without an eavesdropper ({\em i.e.,} when the eavesdropper is jamming and not listening), using a stochastic codebook does not offer any advantage over a conventional codebook, but it does not hurt either, i.e., the receiver still reliably decodes the transmitted codeword \cite{Csiszar78}.

\subsection{CSI Model}
Given the signal framework introduced above, we are ready to discuss
the important issue of CSI.  We have already indicated that Alice
knows $\mathbf{H}_{ba}$ in order to appropriately precode the jamming
and information signals via $\mathbf{T}$ and $\mathbf{T}'$, conceivably obtained by public feedback from Bob after a training phase. At the
receiver side, we will assume that Eve knows the channel from Alice
$\mathbf{H}_{ea}$ and the covariance $\mathbf{K}_e$ of the
interference and noise, and similarly we will assume that Bob knows
$\mathbf{H}_{ba}$ and $\mathbf{K}_b$.  All other CSI at the various
nodes is assumed to be non-informative; the only available information
is that the channels are composed of independent $\mathcal{CN}(0,1)$
random variables.  This implies that when Eve jams Bob, her lack of
information about $\mathbf{H}_{be}$ and the half-duplex constraint
prevents her from detecting the transmitted signal $\mathbf{z}$ and
applying correlated jamming \cite{Basar04}.  Consequently, she will be
led to uniformly distribute her available power over all $N_e$
transmit dimensions, so that $\mathbf{Q}_{be} = \frac{P_e}{N_e}
\mathbf{I}$.  Similarly, when Alice transmits a jamming signal, it
will also be uniformly distributed across the $N_a-d$ available
dimensions: $\mathbf{Q}'_z = \frac{(1-\rho)P_a}{N_a-d} \mathbf{I}$.
While in principle Alice could use her knowledge of $\mathbf{H}_{ba}$
to perform power loading, for
simplicity and robustness we will assume that the power of the information signal is
also uniformly distributed, so that $\mathbf{Q}_z=\frac{\rho P_a}{d}
\mathbf{I}$.

Given the above assumptions,
equations~(\ref{QT})-(\ref{eq:Qb_Qe}) will simplify to
\begin{eqnarray}
\mathbf{Q}_a & = & \frac{\rho P_a}{d}\mathbf{T}\mathbf{T}^H + \eta_a\mathbf{T}'\mathbf{T}'^H \\
\mathbf{K}_b & = & \frac{g_2 P_e}{N_e}\mathbf{H}_{be}\mathbf{H}_{be}^H +\sigma_b^2 \mathbf{I} \\
\mathbf{K}_e & = & g_1\eta_a\mathbf{H}_{ea}\mathbf{T}'\mathbf{T}'^H \mathbf{H}_{ea}^H +\sigma_e^2 \mathbf{I},\;
\end{eqnarray}
where we have defined $\eta_a = \frac{(1-\rho)P_a}{N_a-d}$.

\subsection{Secrecy Rates and Transmit Strategies}
The MIMO secrecy capacity between Alice and Bob is obtained by solving \cite{OggierH08,Shitz09,Wornell09}
\begin{equation}
C_s  = \mathop {\max }\limits_{{\mathbf{Q}}_a  \succeq 0} I\left( {{\mathbf{X}}_a ;{\mathbf{Y}}_b } \right)
- I\left( {{\mathbf{X}}_a ;{\mathbf{Y}}_e } \right) \; ,
\end{equation}
where ${\mathbf{X}}_a, {\mathbf{Y}}_b, {\mathbf{Y}}_e$ are the random
variable counterparts of the realizations $\mathbf{x}_a, \mathbf{y}_a,
\mathbf{y}_e$.  Given the CSI constraints discussed above, such an
optimization cannot be performed since Alice is unaware of the
instantaneous values of all channels and interference covariance
matrices. Consequently, we choose to work with the
lower bound on the MIMO ergodic secrecy capacity based on Gaussian
inputs and uniform power allocation at all transmitters \cite{GoelN08}:
\begin{equation}
\begin{split}
C_s \geq & \mathcal{E}_\mathbf{H}\left\{ {\log}_2{\left| {{\mathbf{I}} + \frac{\rho P_a}{d}{{\mathbf{H}}_{ba}}{\mathbf{T}}{\mathbf{T}}^H{\mathbf{H}}_{ba}^H{{\mathbf{K}}_b^{-1}}} \right|} \right.\\
&{-}\: \left. {\log}_2{\left| {{{\mathbf{I}}} +\frac{g_1 \rho P_a}{d}{{\mathbf{H}}_{ea}}\mathbf{T}{\bf{T}}^H{\mathbf{H}}_{ea}^H}{{\mathbf{K}}_e^{-1}} \right|}  \right\} \; ,
\label{eq:payoff_E}
\end{split}
\end{equation}
where we define ${\mathbf{H}} \triangleq \left\{{{\mathbf{H}}_{ba},
{\mathbf{H}}_{be}, {\mathbf{H}}_{ea} } \right\}$. This serves as a reasonable metric to assess the relative security of the link and to explain the behavior of the players. Recall that
we assume Alice has instantaneous CSI for the link to Bob and only
statistical CSI for Eve, and the achievability of an ergodic secrecy
rate for such a scenario was shown in \cite{Ulukus07}.
Using ergodic secrecy as
the utility function for the game between Alice and Eve implies that a
large number of channel realizations will occur intermediate to any
changes in their strategy.  That is, the physical layer parameters are
changing faster than higher ({\em e.g.,} application) layer functions
that determine the user's strategy.  Thus, the expectation is taken
over all channel matrices (including $\mathbf{H}_{ba}$), which in
turn provides Alice and Eve with a common objective function, since
neither possesses the complete knowledge of ${\mathbf{H}}$ that is
needed to compute the instantaneous MIMO secrecy rate.

Eve must decide whether to eavesdrop or jam with an arbitrary fraction
of her transmit power. Alice's options include determining how many
spatial dimensions are to be used for data and artificial
interference (if any), and the appropriate fraction $\rho$ that
determines the transmit power allocated to them.  As described in
\cite{GoelN08,Mukherjee09,Zhou09,Hong_TWC11,MILCOM10}, there are several
options available to Alice for choosing $\rho$ and $d$ depending upon
the accuracy of her CSI, ranging from an exhaustive search for optimal
values to lower-complexity approaches based on fixed-rate assumptions.
Numerical results from this previous work have indicated that the
achievable secrecy rate is not very sensitive to these parameters, and
good performance can be obtained for a wide range of reasonable
values.  The general approach of this paper is applicable to
essentially any value for $\rho$ and $d$, although the specific
results we present in the simulation section use a fixed value for $d$
and find the optimal value for $\rho$ based on $d$ under the
assumption that the eavesdropper is in fact eavesdropping, and not
jamming.

In Section~\ref{sec:strats} we show that it is sufficient to consider
a set of two strategies for both players without any loss in
optimality.  In particular, we show that Alice need only consider the
options of either transmitting the information signal with full power,
or devoting an appropriate amount of power and signal dimensions to a
jamming signal. On the other hand, Eve's only reasonable strategies
are to either eavesdrop passively or jam Bob with all her available
transmit power.

We will denote Eve's set of possible actions as $\{E,J\}$ to indicate
either ``Eavesdropping'' or ``Jamming,'' while Alice's will be
expressed as $\{F,A\}$ to indicate ``Full-power'' devoted to the
information signal, or a non-zero fraction of the power allocated to
``Artificial interference.'' The secrecy rates that result from the
resulting four possible scenarios will be denoted by $R_{ik}$, where
$i\in \{F,A\}$ and $k\in\{E,J\}$.


Assuming Gaussian inputs $\mathbf{z}$ and $\mathbf{z'}$, the MIMO
secrecy rate between Alice and Bob when Eve is in eavesdropping mode
is
\begin{equation}
\begin{split}
{R_{iE}} =& \mathcal{E}_\mathbf{H}\left\{ {\log_2}{\left| {{{\mathbf{I}}} + \frac{\rho P_a}{d \sigma_b^2}
{{\mathbf{H}}_{ba}}{\mathbf{T}}{\mathbf{T}}^H{\mathbf{H}}_{ba}^H} \right|} \right.\\
&{-}\:\left.{\log_2}{\left| {{{\mathbf{I}}} + \frac{g_1\rho P_a}{d}{{\mathbf{H}}_{ea}}\mathbf{T}{\bf{T}}^H{\mathbf{H}}_{ea}^H}{{\mathbf{K}}_e^{-1}} \right|}  \right\} \; , \label{eq:payoff_E2}
\end{split}
\end{equation}
whereas the secrecy rate when Eve is jamming reduces to
\begin{equation}
{R_{iJ}} = \mathcal{E}_\mathbf{H}\left\{ {\log_2}{\left| {{\mathbf{I}} + \frac{\rho P_a}{d}
{{\mathbf{H}}_{ba}}\mathbf{T}{\mathbf{T}}^H{\mathbf{H}}_{ba}^H}{{\mathbf{K}}_b^{-1}} \right|} \right\},\label{eq:payoff_J}
\end{equation}
where $i = F,A$ denotes the transmission strategies available to
Alice.  We refer to~(\ref{eq:payoff_J}) as a secrecy rate even though
there is technically no eavesdropper, since Eve's mutual information
is identically zero and Alice still uses a stochastic encoder
(cf. Sec.~\ref{sec:model}). Therefore, when evaluating the secrecy
rate definition (11) for the case where Eve chooses to jam, the second
term is zero which yields $R_{FJ}$ and $R_{AJ}$ in (\ref{eq:payoff_J})
as the effective secrecy rate.  Recall that the definition of the
secrecy rate is the maximum transmission rate which can be reliably
decoded by Bob while remaining perfectly secret from Eve, which is
still satisfied by the rates in (\ref{eq:payoff_J}).  Note also that
when Alice employs artificial interference, a choice for $\rho$ and $d$ must
be made that holds regardless of Eve's strategy. Therefore, the values
of $\rho$ and $d$ that are numerically computed to maximize $R_{AE}$
in (\ref{eq:payoff_E2}) \cite{GoelN08} remain unchanged for $R_{AJ}$
in (\ref{eq:payoff_J}).  When Alice transmits with full power, then $d
= r$, where $r = \min(N_a,N_b)$, and the precoder $\mathbf{T}$
consists of the right singular vectors of $\mathbf{H}_{ba}$
corresponding to the $r$ largest singular values.

While Alice uses the same type of encoder regardless of Eve's
strategy, achieving the rates
in~(\ref{eq:payoff_E2})-(\ref{eq:payoff_J}) requires adjustments to
the code rate that {\em will} depend on Eve's actions.  For example,
if Alice is transmitting with full power (strategy $F$), the code rate
needed to achieve either $R_{FE}$ or $R_{FJ}$ in~(\ref{eq:payoff_E2})
or~(\ref{eq:payoff_J}) will be different.  Thus, we assume that Alice
can be made aware of Eve's strategy choice, for example through
feedback from Bob, in order to make such adjustments\footnote{Based on
such feedback, Alice could also in principle switch from a stochastic
encoder to a more standard non-secure code if she discovers that Eve
is jamming and not eavesdropping.  In either case, the rate
expressions in~(\ref{eq:payoff_E2})-(\ref{eq:payoff_J}) will be
valid.}.  Such behavior is not limited to just Alice and Bob; Eve also
makes adjustments based on Alice's choice of strategy.  In particular,
when Eve is eavesdropping, her method of decoding Alice's signal will
depend on whether or not Alice is transmitting artificial interference.  We
do not consider adjustments such as these as part of Alice or Eve's
strategy {\em per se}, which in our game theory framework is
restricted to the decision of whether or not to use artificial
interference.  We assume that minor adaptations to the coding or
decoding algorithm for Alice and Eve occur relatively quickly, and
that any resulting transients are negligible due to our use of ergodic
secrecy rate as the utility function.  The more interesting question
is whether or not Alice and Eve decide to change strategies based on
the actions of the other is addressed in Section~\ref{sec:extensive}.

In the game-theoretic analysis of the next two sections, we will utilize
the following general properties of the MIMO wiretap channel:
\begin{enumerate}\label{list:conditions}
 \item[(\emph{P}1)] $R_{FE}\leq R_{AE}$
 \item[(\emph{P}2)] $R_{AJ}\leq R_{FJ}$
\end{enumerate}
The validity of (\emph{P}2) is obvious; if Alice employs artificial
interference, it reduces the power allocated to the information
signal, which in turn can only decrease the mutual information at Bob.
Since Eve is jamming, her mutual information is zero regardless of
Alice's strategy, so $R_{AJ}$ can never be larger than $R_{FJ}$.  The
validity of (\emph{P}1) can be established by recalling that Alice
chooses a value for $\rho$ that maximizes $R_{AE}$, assuming Eve is
eavesdropping.  Since $\rho=1$ is an available option and corresponds
to $R_{FE}$, Alice can do no worse than $R_{FE}$ in choosing the
optimal $\rho$ for strategy $R_{AE}$.

\section{STRATEGIC WIRETAP GAME}\label{sec:strats}

In this section we construct the zero-sum model of the proposed
wiretap game.  We define the payoff to Alice as the achievable MIMO
secrecy rate between her and Bob. Modeling the strategic interactions
between Alice and Eve as a strictly competitive game leads to a
zero-sum formulation, where Alice tries to maximize her payoff and Eve
attempts to minimize it.

Formally, we can define a compact strategy space $A_i,i=1,2,$ for both
Alice and Eve: Alice has to optimize the pair $(d,\rho)\in A_1$, where
$\rho$ is chosen from the unit interval $[0,1]$ and
$d\in\{1,\ldots,r=\min(N_a,N_b)\}$; and Eve can choose her jamming
power $P_j\in A_2$ from the interval $[0,P_e]$, where zero jamming
power corresponds to the special case of passive eavesdropping. In
other words, each player theoretically has a
continuum of (pure) strategies to choose from, where the
payoff for each combination of strategies is the
corresponding MIMO secrecy rate.  In the following discussion,
let $\left( {d_s^*,\rho _s^*}
\right)$ represent the choice of Alice's parameters that maximizes the
ergodic secrecy rate $R_{AE}$.

The complete set of mixed strategies for player $i$ is the set of Borel probability measures on $A_i$. Let $\Delta_i$ be the set of all probability measures that assign strictly positive mass to every nonempty open subset of $A_i$. The optimal mixed strategy for player $i$ must belong to $\Delta_i$, since any pure strategies that are assigned zero probability in equilibrium can be pruned without changing the game outcome. Furthermore, as in the case of finite games, the subset of pure strategies included in the optimal mixed strategy must be \emph{best responses} to particular actions of the opponent \cite{Petrosjan}.
Consider Alice: when Eve chooses the action of eavesdropping, $\left( {d_s^*,\rho _s^*} \right)$ is Alice's corresponding best response pure strategy since by definition it offers a payoff at least as great as \emph{any} other possible choice of $\left(d,\rho\right)$ [cf. (\emph{P}1)]. Similarly, when Eve chooses to jam with any arbitrary power, Alice's best response pure strategy is $\left( {d = r,\rho  = 1} \right)$ [cf. (\emph{P}2)]. Therefore, these two pure strategies are Alice's best responses for any possible action by Eve, and it is sufficient to consider them alone in the computation of the optimal mixed strategy since all other pure strategies are assigned zero probability. A similar argument holds for Eve with her corresponding best responses of $P_j=0$ and $P_j=P_e$.

Therefore, it is sufficient to consider the following strategy sets
$\mathcal{X},\mathcal{Y}$ for the players: Alice chooses between
transmitting with full power for data (\emph{F}) or devoting an appropriate fraction of
power to jam Eve (\emph{A}), described as $\mathcal{X} = \left\{ {F,A}
\right\}$. Eve must decide between eavesdropping
(\emph{E}) or jamming Bob with full power $P_e$ (\emph{J}) at every channel use, represented
by $\mathcal{Y} = \left\{ {E,J} \right\}$.

\subsection{Pure-strategy Equilibria}

The strategic form of the game where Alice and Eve move simultaneously without observing each other's actions can be represented by the $2 \times
2$ payoff matrix $\mathbf{R}$ in Table~\ref{table:game}.  Our first
result establishes the existence of Nash equilibria for the strategic
game.

\emph{Proposition 1}: For an arbitrary set of antenna array sizes,
transmit powers and channel gain parameters, the following
unique pure-strategy saddle-points or Nash Equilibria (NE)
$\left( {x^* ,y^* } \right)$ exist in the proposed MIMO
wiretap game:
\begin{subnumcases}{{\mathbf{R}}\left( {x^* ,y^* } \right) =}\label{eq:Prop1}
{R_{AE} } & ${{\text{if}}}\quad R_{AE}  \leq R_{AJ}$\\
{R_{FJ} } & ${{\text{if}}}\quad R_{FJ}  \leq R_{FE}$.
\end{subnumcases}
\emph{Proof}: Of the 24 possible orderings of the four rate outcomes,
only six satisfy both conditions (\emph{P}1)-(\emph{P}2) of the
previous section.  Furthermore, it is easy to check that only two of
these six mutually exclusive outcomes results in a pure NE.  If
$R_{AE} \leq R_{AJ}$, then assumptions (\emph{P}1) and
(\emph{P}2) imply the following rate ordering
\begin{equation}
R_{FJ}  \geq R_{AJ}  \geq \underbrace {R_{AE} }_{NE} \geq R_{FE} \label{eq:PureNEorder1} \; .
\end{equation}
In this case, $R_{AE}$ represents an NE since neither Alice nor Eve
can improve their respective payoffs by switching strategies; {\em
i.e.,} the secrecy rate will decrease if Alice chooses to transmit the
information signal with full power, and the secrecy rate will increase
if Eve decides to jam.  Similarly, when $R_{FJ} \le R_{FE}$, then
(\emph{P}1)-(\emph{P}2) result in the rate ordering
\begin{equation}
R_{AE}  \geq R_{FE}  \geq \underbrace {R_{FJ} }_{NE} \geq R_{AJ} \label{eq:PureNEorder2} \; ,
\end{equation}
and $R_{FJ}$ will be the mutual best response for both players.
Evidently only one such ordering can be true for a given wiretap game
scenario.$ \blacksquare$

\subsection{Mixed-strategy Equilibria}
Proposition 1 establishes that there is no single pure strategy choice
that is always optimal for either player if the inequalities
in~(\ref{eq:PureNEorder1})-(\ref{eq:PureNEorder2}) are not
satisfied. This occurs in four of the six valid rate orderings of the
entries of $\mathbf{R}$ that satisfy conditions (\emph{P}1)-(\emph{P}2).
Therefore, since the minimax theorem guarantees that any finite
zero-sum game has a saddle-point in randomized strategies
\cite{Myerson}, in such scenarios Alice and Eve should randomize over
$\mathcal{X} \times \mathcal{Y}$; that is, they should adopt mixed
strategies.

\begin{figure}[htp]
\centering
\includegraphics[width=\linewidth]{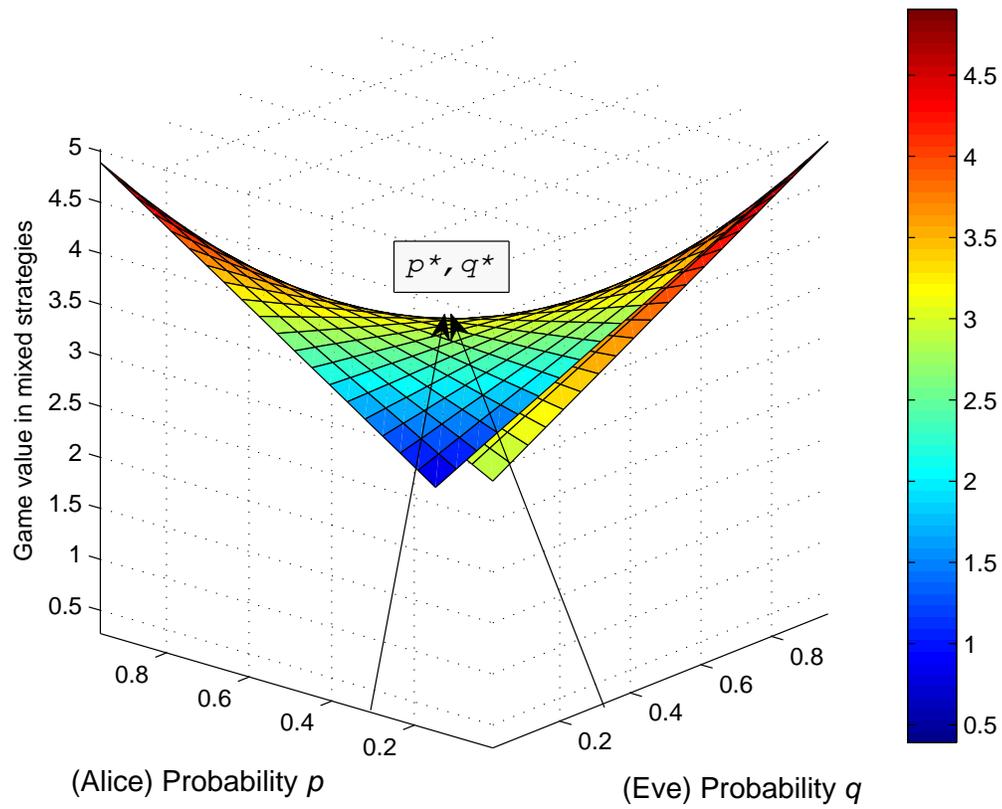}
\caption{Game value in mixed strategies as the mixing probabilities at Alice and Eve are varied, $N_a=5,N_b=3,N_e=4,d=2,$ $P_a=P_e=20$dB, and $g_1=1.1,g_2=0.9$.}
\label{fig_MixedStrats_3D}
\end{figure}

Let $\mathbf{p}=(p,1-p)$ and $\mathbf{q}=(q,1-q)$, $0\leq p,q\leq 1,$
represent the probabilities with which Alice and Eve randomize over
their strategy sets $\mathcal{X}=\left\{ {F,A} \right\}$ and
$\mathcal{Y}=\left\{ {E,J} \right\}$, respectively. In other words,
Alice plays {$x=F$} with probability $p$, while Eve plays {$y=E$} with
probability $q$. Alice obtains her optimal strategy by solving
\begin{equation}\label{eq:maxmin}
\mathop {\max }\limits_p \mathop {\min }\limits_q {\mathbf{p}}^T {\mathbf{Rq}},
\end{equation}
while Eve optimizes the corresponding minimax problem. For the payoff
matrix $\mathbf{R}$ in Table~\ref{table:game}, the optimal mixed
strategies and unique NE value $v$ of the game can be easily derived as \cite{Fudenberg,Myerson}
\begin{subequations}\label{eq:mixed}
\begin{align}
  \left( {p^*,1 - p^*} \right) &= {{\left( {R_{AJ}  - R_{AE} ,R_{FE}  - R_{FJ} } \right)} \mathord{\left/
 {\vphantom {{\left( {R_{AJ}  - R_{AE} ,R_{FE}  - R_{FJ} } \right)} D}} \right.
 \kern-\nulldelimiterspace} D} \hfill \label{eq:Alicemaximin}\\
  \left( {q^*,1 - q^*} \right) &= {{\left( {R_{AJ}  - R_{FJ} ,R_{FE}  - R_{AE} } \right)} \mathord{\left/
 {\vphantom {{\left( {R_{AJ}  - R_{FJ} ,R_{FE}  - R_{AE} } \right)} D}} \right.
 \kern-\nulldelimiterspace} D} \hfill \label{eq:Eveminimax}\\
  v(p^*,q^*) &= {{\left( {R_{FE} R_{AJ}  - R_{FJ} R_{AE} } \right)} \mathord{\left/
 {\vphantom {{\left( {R_{FE} R_{AJ}  - R_{FJ} R_{AE} } \right)} D}} \right.
 \kern-\nulldelimiterspace} D}, 
\end{align}
\end{subequations}
where $D = R_{FE} + R_{AJ} - R_{FJ} - R_{AE}$. The mixed NE above is unique according to the classic properties of finite matrix games \cite{Fudenberg}, since the optimization in \eqref{eq:maxmin} has a unique solution.  A graphical
illustration of the saddle-point in mixed strategies as $p$ and $q$
are varied for a specific wiretap channel is shown in
Fig.~\ref{fig_MixedStrats_3D}. For the specified parameters
$N_a=5,N_b=3,N_e=4,d=2,$ $P_a=P_e=20$dB, $g_1=1.1,g_2=0.9$, the
rate ordering turns out to be $R_{AE}=5.04> R_{FJ}=5.02>{R_{AJ}=2.85 }
> R_{FE}=0$, which results in a mixed NE with optimal mixing
probabilities $(p*=0.307,q*=0.294)$ and value $v=3.45$. Alice's bias
towards playing $x=A$ more frequently is expected since that
guarantees a secrecy rate of at least 2.85, whereas playing $x=F$
risks a worst-case payoff of zero. Eve is privy to Alice's reasoning
and is therefore biased towards playing $y=J$ more frequently since
she prefers a game value close to $R_{AJ}$.

The \emph{repeated} wiretap game is a more sophisticated strategic game model in which Alice and Eve play against each other repeatedly over multiple stages in time. At each stage, the set of player strategies and payoff function representation is identical to the single-stage zero-sum game $\mathbf{R}$ in Table~\ref{table:game}. In our context, the single-stage game can be considered to represent the transmission of a single codeword, with the repeated game spanning the successive transmission of multiple codewords. Let the payoff to Alice at stage $k$ be denoted as ${R\left[ k \right]}$. Under the \emph{standard repeated game model} \cite{Myerson}, the payoffs are accrued after each stage, and both players have perfect information of the adversary's moves. If the game is repeated over an infinite time horizon, the cumulative payoff (of Alice) over the duration of the game is given by
\begin{equation}
{R_p} = \left( {1 - \delta } \right)\sum\limits_{k = 0}^\infty  {{\delta ^k}R\left[ k \right]}
\end{equation}
where the discounting factor $\delta$, $0 \leq \delta  < 1$, ensures that $R_p$ is finite. Unlike general nonzero-sum repeated games where players can improve payoffs via cooperation over time \cite{Liu09}, the strictly competitive nature of the zero-sum wiretap game results in Alice and Eve repeatedly playing their single-stage game NE strategies. For example, it is clear that Eve minimizes $R_p$ by minimizing $R\left[ k \right]$ at each stage $k$, which is achieved by playing as dictated by Proposition 1 or \eqref{eq:mixed} at each stage.
If the game is played over a finite number of stages instead, the players will continue to play their single-stage game NE strategies by the same argument. The concepts developed in Sec.~\ref{sec:imperfectinfo} are applicable to the more involved repeated game scenario where Alice and Eve have imperfect observations of each other's actions.

\section{EXTENSIVE FORM WIRETAP GAME}\label{sec:extensive}

Given the strategic game analysis of the previous section, we can now
proceed to analyze the actions of a given player in response to the
opponent's strategy.  Here, one player is assumed to move first,
followed by the opponent's response, which can then lead to a strategy (and code rate)
change for the first player, and so on.  Accordingly, in this section
we examine the sequential or \emph{extensive form} of the MIMO wiretap
game, which is also known as a Stackelberg game. The standard analysis of a Stackelberg game is to cast it as a dynamic or extensive-form game and elicit equilibria based on backward induction \cite{Fudenberg}.  We begin with the
worst-case scenario where Alice moves first by either playing \emph{F}
or \emph{A}, which is observed by Eve who responds accordingly. It is
convenient to represent the sequential nature of an extensive-form
game with a rooted tree or directed graph, as shown in
Fig. \ref{fig:extensive}. The payoffs for Alice are shown at each
terminal node, while the corresponding payoffs for Eve are omitted for
clarity due to the zero-sum assumption. In this section, we explore
extensive-form games with and without perfect information, and the
variety of equilibrium solution concepts available for them.

\subsection{Perfect Information}\label{sec:perfectinfo}
\begin{figure}[t]
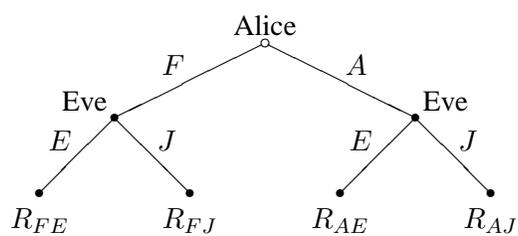

\hspace*{\fill}
\begin{egame}(600,380)
\putbranch(300,240)(2,1){200}
\iib{Alice}{$F$}{$A$}
\putbranch(100,140)(1,1){100}
\iib{Eve}[l]{$E$}{$J$}[$R_{FE}$][$R_{FJ}$]
\putbranch(500,140)(1,1){100}
\iib{Eve}[r]{$E$}{$J$}[$R_{AE}$][$R_{AJ}$]
\end{egame}
\hspace*{\fill}
\caption[]{Extensive form game tree with perfect information $\Gamma^{e,1}$ where Alice moves first and Eve moves second.}\label{fig:extensive}
\end{figure}

Assuming that Eve can distinguish which move was adopted by Alice,
and furthermore determine the exact jamming power $(1-\rho) P_a$ if
she is being jammed by Alice, then the extensive game is classified as
one of \emph{perfect information}. In the sequel, we will make use of
the notions of an \emph{information state} and a \emph{subgame}. A
player's information state represents the node(s) on the decision tree
at which she must make a move conditioned on her knowledge of the
previous move of the opponent. For the case of perfect information in
Fig.~\ref{fig:extensive}, Alice has a single information state, while
Eve has two information states (each with a single node) based on
Alice's choice, since she has perfect knowledge of Alice's move. A
subgame is a subset (subgraph) of a game that starts from an
information state with a single node, contains all of that
node's successors in the tree, and contains all or none of the nodes
in each information state \cite{Myerson}.

%

Next, we analyze \emph{subgame-perfect equilibria} (SPE) of the
extensive game, which are a more refined form of NE that eliminate
irrational choices within subgames \cite{Fudenberg,Myerson}.  It is
well known that in extensive games with perfect information, a
sequential equilibrium in pure strategies is guaranteed to exist
\cite[Theorem 4.7]{Myerson}. The equilibrium strategies can be
obtained by a process of backward induction on the extensive game
tree, as shown below.

\emph{Proposition 2}: In the extensive form wiretap game $\Gamma^{e,1}$ with perfect information where Alice moves first, the unique subgame-perfect equilibrium rate with pure strategies is determined by the following:
\begin{subnumcases}{\text{SPE}\left(\Gamma^{e,1}\right)=}
R_{A,E}  & ${\text{if }}\; R_{AE}  \leq R_{AJ}$\nonumber\\
R_{F,J}  & ${\text{if }}\; R_{FJ}  \leq R_{FE}$\nonumber\\
\max\left[{R_{FE},R_{AJ} }\right] & $R_{FE}\leq R_{FJ}, {R_{AJ} \leq R_{AE}}$ \nonumber
\end{subnumcases}
\emph{Proof}: The extensive game tree for this problem, depicted in
Fig.~\ref{fig:extensive}, is comprised of three subgames: the two
subgames at Eve's decision nodes, and the game itself with Alice's
decision node as the root. Consider the scenario ${R_{FE} \leq R_{FJ}
\text{ and }} {R_{AJ} \leq R_{AE}}$.  Under this assumption, Eve
always plays $E$ in the lower-left subgame of
Fig.~\ref{fig:extensive}, whereas Eve picks $J$ in the lower-right
subgame. By backward induction, Alice then chooses the larger of
$\left[{R_{FE},R_{AJ} }\right]$ at her decision node. The other two
SPE outcomes can be established in a similar manner. $\blacksquare$

\emph{Proposition 3}: The extensive form game $\Gamma^{e,2}$ with
perfect information where Eve moves first and Alice moves second has
the following subgame-perfect equilibrium rate outcome and corresponding strategies:
\begin{equation}
\text{SPE}\left(\Gamma^{e,2}\right)=\min\left[R_{FJ},R_{AE}\right] \; .
\end{equation}
\emph{Proof}: The extensive game tree for this scenario is depicted in
Fig.~\ref{fig:extensive2}, and is comprised of three subgames: the two
subgames at Alice's decision nodes, and the game itself with Eve's
decision node as the root. Based on properties
(\emph{P}1)-(\emph{P}2), Alice always plays $A$ in the lower-left
subgame and $F$ in the lower-right subgame. By backward induction, Eve
then chooses the action corresponding to the smaller payoff between
$\left[{R_{AE},R_{FJ} }\right]$ at her decision node.  $\blacksquare$

Note that in the scenario where Alice moves first, she chooses
her coding parameters based on the assumption that Eve acts rationally
and adopts the equilibrium strategy in Proposition 2.  We see from
both propositions that, when conditions for one of the pure-strategy
NEs hold, the outcome of both $\Gamma^{e,1}$ and $\Gamma^{e,2}$ will
be the corresponding NE.  This is also true of an extensive game with
more than 2 stages; if an NE exists, the overall SPE outcome will be
composed of repetitions of this constant result. 

\subsection{Imperfect Information}\label{sec:imperfectinfo}

We now consider extensive wiretap games with imperfect information, where the player moving second has an imperfect estimate of the prior move made by her opponent. Let $\Gamma^{e,3}_f$ and $\Gamma^{e,4}_f$ denote the games where Alice and Eve move first, respectively.
The game tree
representation of $\Gamma^{e,3}_f$ can be drawn by connecting the
decision nodes of Eve in Fig.~\ref{fig:extensive} to indicate her
inability to correctly determine Alice's move in the initial phase of
the game.  Thus, in this case, Eve effectively only possesses a single
information state.  While no player has an incentive to randomize in
the game with perfect information in Section~\ref{sec:perfectinfo},
mixed strategies enter the discussion when the game is changed to one
of imperfect information. The subgame perfect equilibrium solution is
generally unsatisfactory for such games, since the only valid subgame
in this case is the entire game $\Gamma^{e,3}_f$ itself. Therefore,
\emph{sequential equilibrium} is a stronger solution concept better
suited for extensive games of imperfect information.

An extreme case of imperfect information in $\Gamma^{e,3}_f$ is the scenario where it is common knowledge at all
nodes that Eve is \emph{completely unable} to determine what move was made by
Alice in the first stage of the game. Let Eve then assign the \emph{a priori}
probabilities $\left({\alpha},{1-\alpha}\right)$ to Alice's moves over
$\left\{ {F,A} \right\}$ for some $\rho$ and $d$, while Eve herself
randomizes over $\{E,J\}$ with probabilities
$\left({\gamma},{1-\gamma}\right)$. Therefore, Eve's left and right decision nodes are reached with probability $\alpha$ and $\left(1-\alpha \right)$, respectively. There are three possible supports for Eve's moves at her information state: pure strategies $\{E\}$ or $\{J\}$ exclusively, or randomizing over $\{E,J\}$. In the general scenario where Eve randomizes over $\{E,J\}$ with probabilities $\left({\gamma},{1-\gamma}\right)$, her expected payoff can be expressed as
\[
   - \alpha \left[ {\gamma R_{FE}  + \left( {1 - \gamma } \right)R_{FJ} } \right] +   \left( {\alpha  - 1} \right)\left[ {\gamma R_{AE}  + \left( {1 - \gamma } \right)R_{AJ} } \right].
\]
Using a probabilistic version of backward induction, it is straightforward to compute the sequential equilibrium of $\Gamma^{e,3}_f$, which in fact turns out to be identical to the mixed-strategy NE in (\ref{eq:mixed}). A similar argument holds for $\Gamma^{e,4}_f$ with no information at Alice, which arises if no feedback is available from Bob.

It is much more reasonable to assume that the player moving second is able to form some estimate of her opponent's move, known as the \emph{belief} vector \cite{Myerson}. An example of how such a scenario may play out is described here. Consider the game $\Gamma^{e,4}_f$, where Alice's belief vector represents the posterior probabilities of Eve having played \{E\} and \{J\} in the first stage. Assume that Bob collects $M$ signal samples and provides Alice with an inference of Eve's move via an error-free public feedback channel. The competing hypotheses at Bob are
\begin{equation}
\begin{array}{*{20}{c}}
  {{\mathcal{H}_0}:}&{{{\mathbf{y}}_b}\left[ n \right] = {{\mathbf{H}}_{ba}}{{\mathbf{x}}_a}\left[ n \right] + {{\mathbf{n}}_b}\left[ n \right]} \\
  {{\mathcal{H}_1}:}&{{{\mathbf{y}}_b}\left[ n \right] = {{\mathbf{H}}_{ba}}{{\mathbf{x}}_a}\left[ n \right] + \sqrt {{g_2}} {{\mathbf{H}}_{be}}{{\mathbf{x}}_e}\left[ n \right] + {{\mathbf{n}}_b}\left[ n \right]\:,}
\end{array}
\end{equation}
for $n=0,\ldots,M-1,$ where the null hypothesis ${\mathcal{H}_0}$ corresponds to Eve listening passively and the alternative hypothesis ${\mathcal{H}_1}$ is that she is jamming Bob. Here, the channels are assumed to be constant over the sensing interval \cite{Gazor10} and known to Bob since he possesses local CSI. Aggregating the samples into a $(N_b \times M)$ matrix ${{{\mathbf{Y}}_b} = \left[ {\begin{array}{*{20}{c}}
  {{{\mathbf{y}}_b}\left[ 0 \right]}& \ldots &{{{\mathbf{y}}_b}\left[ {M - 1} \right]}
\end{array}} \right]}$, we observe that ${{{\mathbf{Y}}_b} \sim \mathcal{CN}\left( {{\mathbf{0}},{{\mathbf{Z}}_0}} \right)}$ under ${\mathcal{H}_0}$ and ${{{\mathbf{Y}}_b} \sim \mathcal{CN}\left( {{\mathbf{0}},{{\mathbf{Z}}_0}+{{\mathbf{Z}}_1}} \right)}$ under ${\mathcal{H}_1}$, where
\begin{align*}
{{\mathbf{Z}}_0}  \triangleq & {{\mathbf{H}}_{ba}}{{\mathbf{Q}}_a}{\mathbf{H}}_{ba}^H + \sigma _b^2{\mathbf{I}}\\
{{\mathbf{Z}}_1} \triangleq &\left( \frac{P_e}{N_e} \right){{\mathbf{H}}_{be}}{\mathbf{H}}_{be}^H.
 \end{align*}

Assuming that Bob employs a minimum probability of error (MPE) detector \cite{KayVolII}, the hypothesis test is
\begin{equation}\label{eq:BobMPEtest}
\frac{{f\left( {{{\mathbf{Y}}_b}|{\mathcal{H}_1}} \right)}}{{f\left( {{{\mathbf{Y}}_b}|{\mathcal{H}_0}} \right)}} \mathop \gtrless \limits_{{{\mathcal{H}}_0}}^{{{\mathcal{H}}_1}} \frac{{\Pr\left( {{\mathcal{H}_1}} \right)}}{{\Pr\left( {{\mathcal{H}_0}} \right)}} = \eta
\end{equation}
where $\Pr\left( {{\mathcal{H}_1}} \right)$ and $\Pr\left( {{\mathcal{H}_0}} \right)$ are prior probabilities assigned to the hypotheses by Bob. A worst-case assumption for the prior probabilities is given by Eve's minimax mixing probabilities in \eqref{eq:Eveminimax}. Taking the logarithm on both sides of \eqref{eq:BobMPEtest} and inserting the appropriate densities
\begin{align}
f\left( {{{\mathbf{Y}}_b}|{H_1}} \right) =& {\pi ^{ - M{N_b}}}{{\left| {{{\mathbf{Z}}_0} + {{\mathbf{Z}}_1}} \right|}^M}\nonumber\\
&{\times}\: \exp \left( { - \Tr\left( {{{\left( {{{\mathbf{Z}}_0} + {{\mathbf{Z}}_1}} \right)}^{ - 1}}{{\mathbf{Y}}_b}{\mathbf{Y}}_b^H} \right)} \right)\\
f\left( {{{\mathbf{Y}}_b}|{H_0}} \right) &= {\pi ^{ - M{N_b}}}{{\left| {{{\mathbf{Z}}_0}} \right|}^M}\exp \left( { - \Tr\left( {{\mathbf{Z}}_0^{ - 1}{{\mathbf{Y}}_b}{\mathbf{Y}}_b^H} \right)} \right),
\end{align}
after some manipulations we obtain the test
\begin{equation}\label{eq:BobMPEtest}
\Tr\left( {\left( {{\mathbf{Z}}_0^{ - 1} - {{\left( {{{\mathbf{Z}}_0} + {{\mathbf{Z}}_1}} \right)}^{ - 1}}} \right){{\mathbf{Y}}_b}{\mathbf{Y}}_b^H} \right) \mathop \gtrless \limits_{{{\mathcal{H}}_0}}^{{{\mathcal{H}}_1}} \eta '
\end{equation}
where $\eta ' = \ln \left( \eta  \right) + {\left| {{{\mathbf{Z}}_0} + {{\mathbf{Z}}_1}} \right|^M} + {\left| {{{\mathbf{Z}}_0}} \right|^M}$.

Finally, Alice determines her best response based on the posterior probabilities (beliefs) of the hypotheses, which is the definition of a sequentially rational strategy \cite{Myerson}. 
The requisite posterior probabilities are ${\alpha _i} \triangleq \Pr \left\{ {{\mathcal{H}_i}|{{\mathbf{Y}}_b}} \right\} = {{f\left( {{{\mathbf{Y}}_b}|{\mathcal{H}_i}} \right)\Pr \left\{ {{\mathcal{H}_i}} \right\}} \mathord{\left/
 {\vphantom {{f\left( {{{\mathbf{Y}}_b}|{\mathcal{H}_i}} \right)\Pr \left\{ {{\mathcal{H}_i}} \right\}} {f\left( {{{\mathbf{Y}}_b}} \right)}}} \right.
 \kern-\nulldelimiterspace} {f\left( {{{\mathbf{Y}}_b}} \right)}}$,$\:i=0,1$, with $\alpha_1=1-\alpha_0$ and ${f\left( {{{\mathbf{Y}}_b}} \right)}={\sum\nolimits_i {\left( {f\left( {{{\mathbf{Y}}_b}|{\mathcal{H}_i}} \right)\Pr \left\{ {{\mathcal{H}_i}} \right\}} \right)} }$. At equilibrium, Alice has by definition no incentive to switch actions, which implies that her expected payoffs are the same. Since her expected payoff if she plays $\{F\}$ is ${\alpha _0}{R_{FE}} + \left( {1 - {\alpha _0}} \right){R_{FJ}}$, and ${\alpha _0}{R_{AE}} + \left( {1 - {\alpha _0}} \right){R_{AJ}}$ if she plays $\{A\}$, it follows that Alice's best response is given by
 \begin{equation}
 B{R_A}\left( {{\alpha _0}} \right) = \left\{ {\begin{array}{*{20}{c}}
  F&{\text{if }}{{{\alpha _0} \leq \frac{\left( {{R_{FJ}} - {R_{AJ}}} \right)}{\left( {{R_{AE}} - {R_{FE}} + {R_{FJ}} - {R_{AJ}}} \right)}}} \\
  A&{{\text{otherwise}.}}
\end{array}} \right.
\end{equation}
On the other hand, since Eve moves first in $\Gamma^{e,4}_f$, she does not have causal knowledge of Alice's beliefs, and therefore continues to play her minimax strategies in \eqref{eq:Eveminimax}.

For the game $\Gamma^{e,3}_f$ where Eve moves second, she forms her beliefs about Alice's move ($\{F\}$ or $\{A\}$) from the binary hypothesis test
\begin{equation}
\begin{array}{l}
\mathcal{H}_0 :{\mathbf{y}}_e  = \sqrt {g_1 } {\mathbf{H}}_{ea} {\mathbf{z}}\left[n\right] + {\mathbf{n}}_e\left[n\right] \\
\mathcal{H}_1 :{\mathbf{y}}_e  = \sqrt {g_1 } {\mathbf{H}}_{ea} {\mathbf{Tz}}\left[n\right] + \sqrt {g_1 }
    {\mathbf{H}}_{ea} {\mathbf{T}}'{\mathbf{z}}'\left[n\right] + {\mathbf{n}}_e\left[n\right]
\end{array}
\end{equation}
for $n=0,\ldots,M-1$.
The $(N_e \times M)$ sample matrix ${{{\mathbf{Y}}_e} = \left[ {\begin{array}{*{20}{c}}
  {{{\mathbf{y}}_e}\left[ 0 \right]}& \ldots &{{{\mathbf{y}}_e}\left[ {M - 1} \right]}
\end{array}} \right]}$ follows the distributions ${{{\mathbf{Y}}_e} \sim \mathcal{CN}\left( {{\mathbf{0}},{{\mathbf{Z}}_0}} \right)}$ under ${\mathcal{H}_0}$ and ${{{\mathbf{Y}}_e} \sim \mathcal{CN}\left( {{\mathbf{0}},{{\mathbf{Z}}_1}} \right)}$ under ${\mathcal{H}_1}$, where
\begin{align*}
{{\mathbf{Z}}_0} \triangleq & \left( \frac{g_1P_a}{N_a} \right){{\mathbf{H}}_{ea}}{\mathbf{H}}_{ea}^H + \sigma_e^2{\mathbf{I}}\\
 {{\mathbf{Z}}_1} \triangleq & \left( \frac{g_1\rho P_a}{N_a} \right){{\mathbf{H}}_{ea}}{\mathbf{T}}{{\mathbf{T}}^H}{\mathbf{H}}_{ea}^H \nonumber\\
 &{+}\: \left( \frac{g_1\left( {1 - \rho } \right)P_a}{N_a} \right){{\mathbf{H}}_{ea}}{\mathbf{T'}}{{{\mathbf{T'}}}^H}{\mathbf{H}}_{ea}^H + \sigma _e^2{\mathbf{I}}.
\end{align*}
The MPE test at Eve thus simplifies to
\begin{equation}\label{eq:BobMPEtest}
\Tr\left( {\left( {{\mathbf{Z}}_0^{ - 1} - {{\left( {{{\mathbf{Z}}_1}} \right)}^{ - 1}}} \right){{\mathbf{Y}}_b}{\mathbf{Y}}_b^H} \right) \mathop \gtrless \limits_{{{\mathcal{H}}_0}}^{{{\mathcal{H}}_1}} \eta '
\end{equation}
where $\eta ' = \ln \left( \eta  \right) + {\left| {{{\mathbf{Z}}_1}} \right|^M} + {\left| {{{\mathbf{Z}}_0}} \right|^M}$, and $\eta$ is the ratio of worst-case prior probabilities based on \eqref{eq:Alicemaximin}. By the equivalence of equilibrium payoffs, Eve's best response based on her computed posterior probabilities $\left(\alpha_0,1-\alpha_0\right)$ is
 \begin{equation}
 B{R_E}\left( {{\alpha _0}} \right) = \left\{ {\begin{array}{*{20}{c}}
  E&{\text{if }{\alpha _0} \leq \frac{\left( {{R_{AE}} - {R_{AJ}}} \right)}{\left( {{R_{AE}} - {R_{FE}} + {R_{FJ}} - {R_{AJ}}} \right)}} \\
  J&{{\text{otherwise}.}}
\end{array}} \right.
\end{equation}
Since Alice has no means of estimating the beliefs possessed by Eve,
Alice plays her maximin strategy as specified by \eqref{eq:Alicemaximin} when she moves first.

\section{SIMULATION RESULTS}\label{sec:sim}
In this section, we present several examples that show the equilibrium
secrecy rate payoffs for various channel and user configurations. All
displayed results are based on the actual numerically computed
secrecy rates with 5000 independent trials per point. NE rates are depicted using a dashed red line where applicable.
In all of the simulations, the noise power was assumed to be the same
for both Bob and Eve: $\sigma_b^2=\sigma_e^2=1$.

\begin{figure}[htp]
\centering
\subfigure[Resulting secrecy rates.]{\label{fig_mix_A}\includegraphics[width=\linewidth]{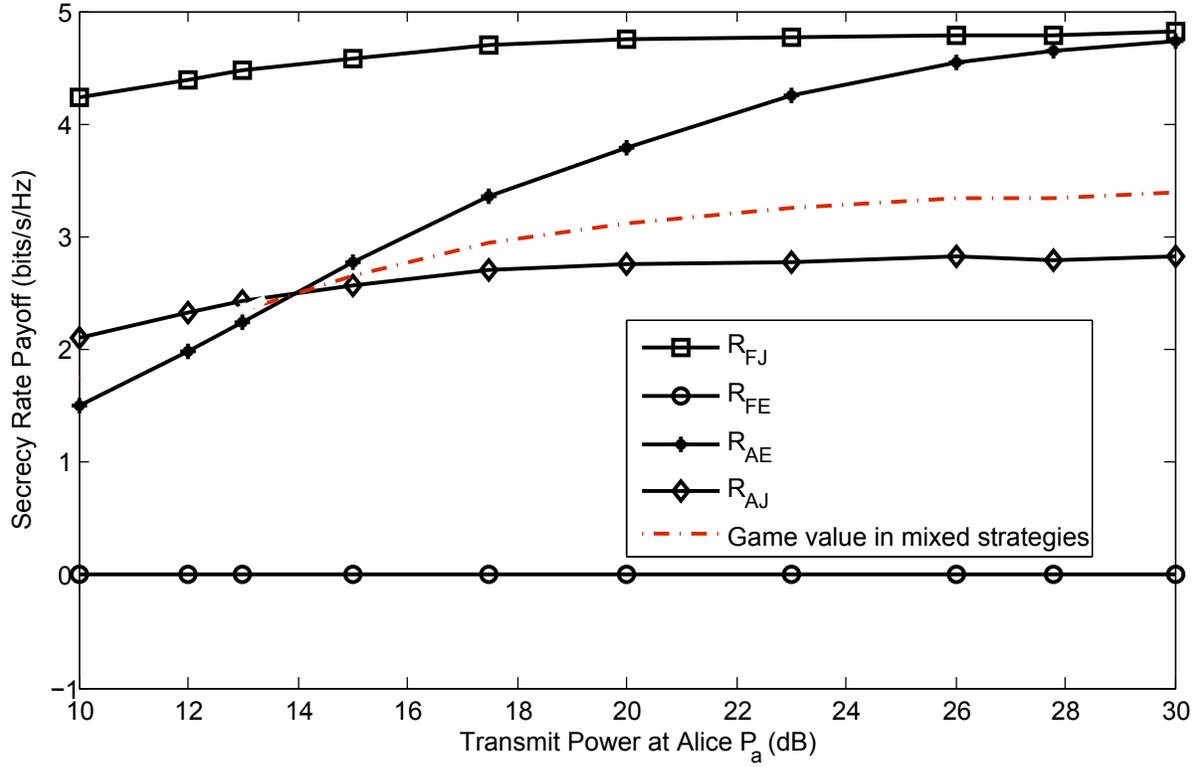}}
\subfigure[Optimal mixing probabilities.]{\label{fig_mix_B}\includegraphics[width=\linewidth]{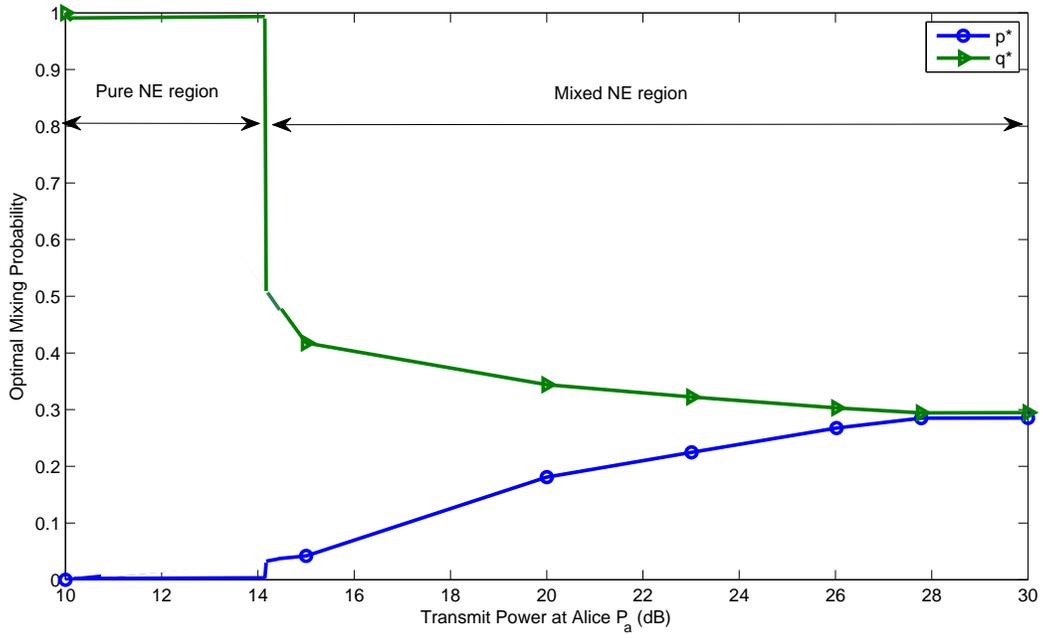}}
\caption{Strategic MIMO wiretap game for $P_e=4P_a$, $N_a=N_e=8,N_b=6,d=4, g_1=1.2, g_2=0.75$ as a function of the transmit power at Alice.}
\label{fig_mix}
\end{figure}

For the strategic game in Fig.~\ref{fig_mix} we set $N_a=N_e=8,N_b=6,$
and Eve's power is larger than Alice's: $P_e = 4P_a$.  The optimal
choice for the signal dimension in this scenario is $d=4$.
Prior to the cross-over, a pure
strategy NE in $R_{AE}$ is the game outcome since the rate ordering is
that of (\ref{eq:PureNEorder1}), whereas after the cross-over it is
optimal for both players to play mixed strategies according to
(\ref{eq:mixed}). In this case, randomizing strategies clearly leads
to better payoffs for the players as Eve's jamming power increases,
compared to adopting a pure strategy. The optimal mixing probabilities
are shown in Fig.~\ref{fig_mix_B} with a clear division between pure
and mixed strategy NE regions.  The pure NE is lost as $P_a$
increases since $R_{AE}$ grows more quickly than $R_{AJ}$.  This is
because increasing $P_a$ under $AE$ both improves Bob's rate and
reduces Eve's rate, since more power is available for both signal and
jamming.  For AJ, increasing $P_a$ can only improve Bob's rate since
Eve is not impacted by the artificial interference (any power devoted
to artificial interference is wasted).

\begin{figure}[htp]
\centering
\includegraphics[width=\linewidth]{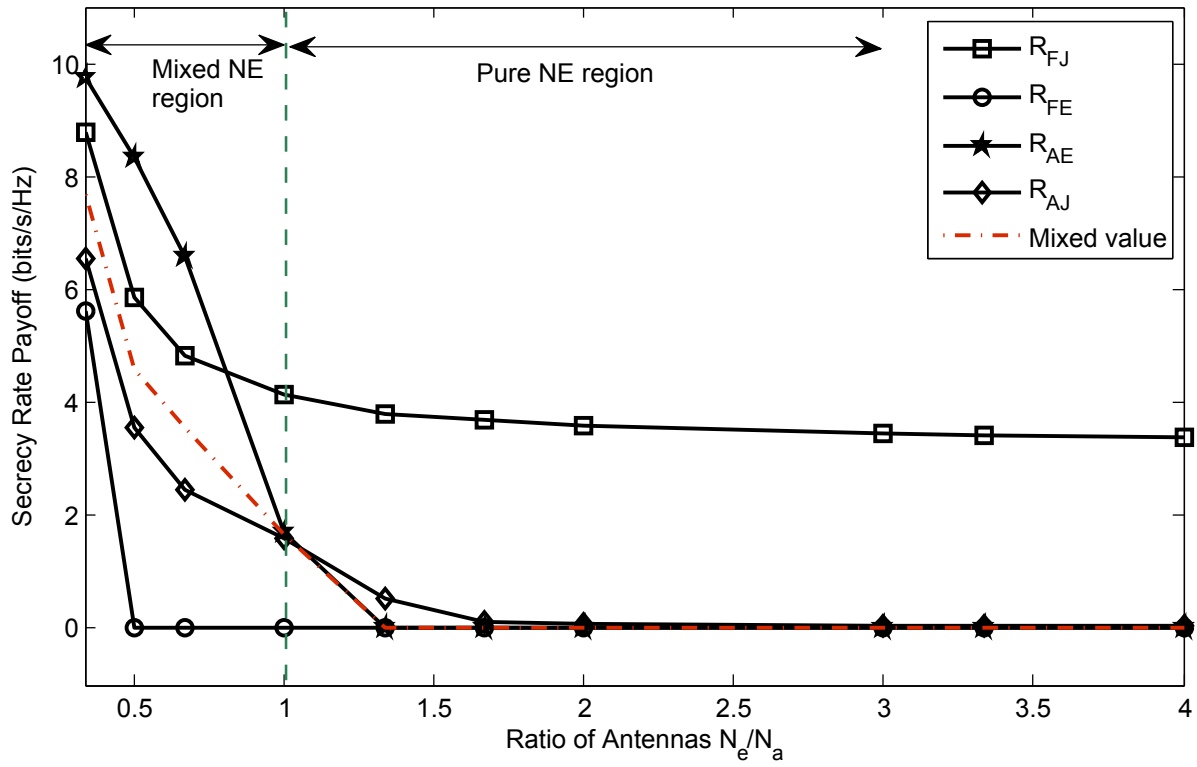}
\caption{Payoff versus antenna ratio $N_e/N_a$ for fixed transmit powers $P_e=P_a=100$ and $N_a=6,N_b=3,d=2,g_1=1.1, g_2=0.9$.}
\label{fig_Antratio}
\end{figure}

For the case of equal transmit powers $P_e = P_a=100$ and parameters
$N_a=6,N_b=3,d=2$, the outcomes of the strategic game as the ratio of
eavesdropper to transmitter antennas varies is shown in
Fig.~\ref{fig_Antratio}. We observe that a similar dichotomy exists
between a pure-strategy saddle-point region and a mixed-strategy
equilibrium in terms of $N_e/N_a$ (with the transition roughly at
$(N_e/N_a)= 1$ marked by the dashed red line).

\begin{figure}[htp]
\centering
\includegraphics[width=\linewidth]{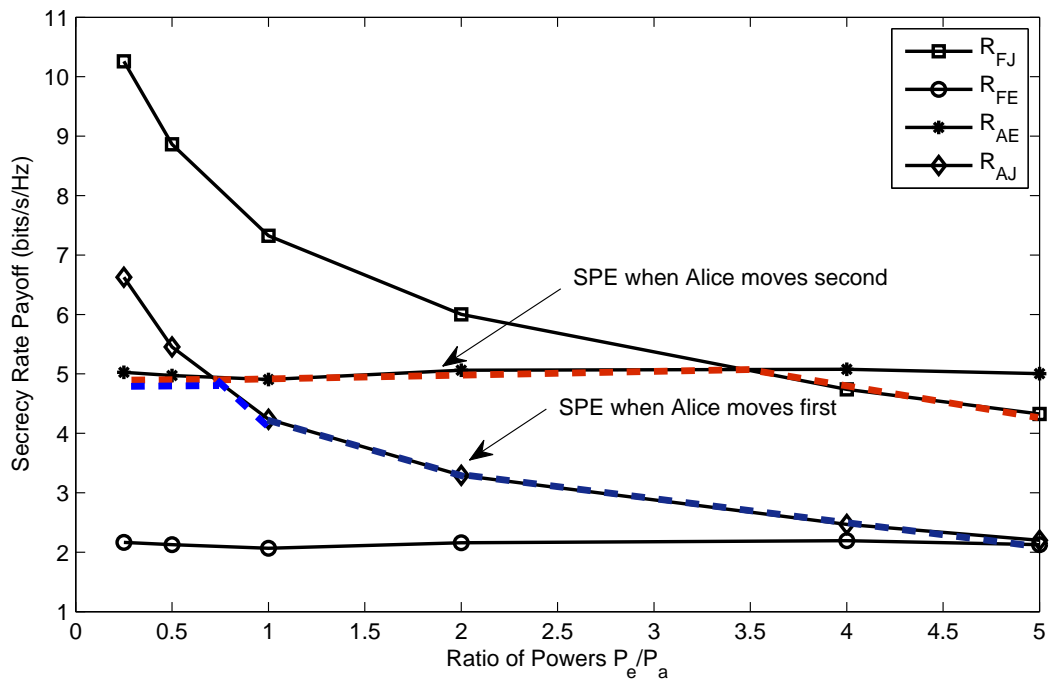}
\caption{Extensive-form games with perfect information, $N_a=N_b=N_e=3,P_a=20dB,d=1, g_1=0.8,g_2=1.1$.}
\label{fig_subgame}
\end{figure}

Next, the SPE outcomes of the two extensive-form games
$\Gamma^{e,1}$ and $\Gamma^{e,2}$ over a range of transmit power
ratios $P_e/P_a$ are shown in Fig.~\ref{fig_subgame}. The red and blue
dashed lines represent the subgame-perfect outcomes of the game where
Alice moves first or second, respectively, as defined in Proposition 2
and Corollary 1. In the extensive form game, Alice could adjust her
transmission parameters ($\rho, d, {\bf T}$, etc.) in addition to her
overall strategy ($A$ or $F$) in response to Eve's move.  For
simplicity, and to allow us to present the main result in a single
figure, we have assumed instead that the transmission parameters are
chosen independently of Eve's actions, as described for the strategic
game. Observe that prior to the crossover point of $R_{AE}$ and
$R_{AJ}$, both equilibria are equal as determined by Proposition~2,
since a pure-strategy NE results. We see that it is always beneficial
for Alice to move second especially as Eve's jamming power increases,
which agrees with intuition.

\begin{figure}[htp]
\centering
\includegraphics[width=\linewidth]{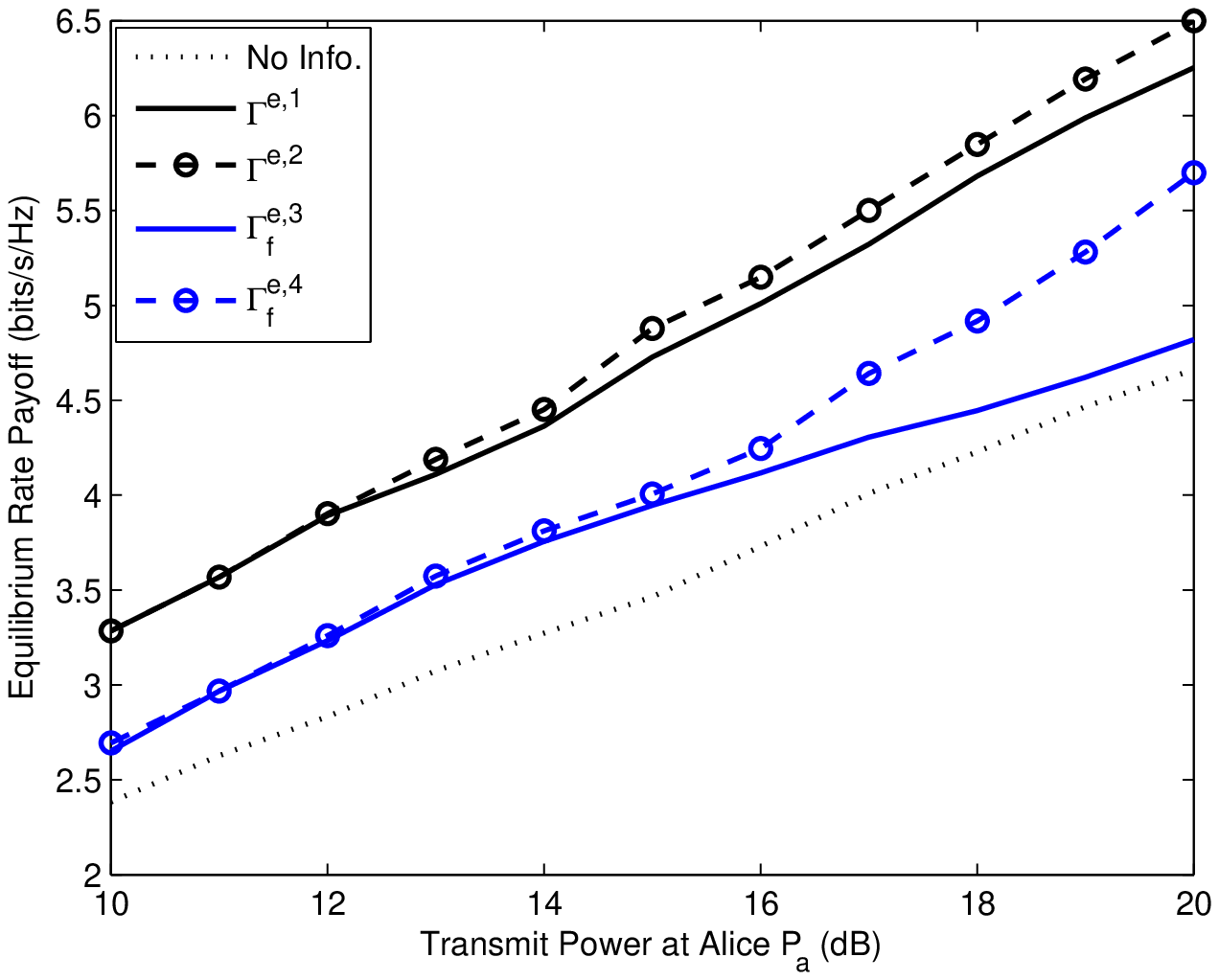}
\caption{Extensive-form games with perfect information, $N_a=N_b=3$,$N_e=2,P_e=2P_a,d=2, g_1=0.8,g_2=1.3$.}
\label{fig_extensiveimperf}
\end{figure}

Finally, in Fig.~\ref{fig_extensiveimperf} we compare the equilibrium outcomes of the extensive-form games with perfect and imperfect information as a function of $P_a$, with $P_e=2P_a$. The no-information lower bound is given by the strategic game mixed-strategy NE. For the given choice of parameters, Alice is not significantly disadvantaged when she moves first $(\Gamma^{e,1})$ in the idealized scenario of perfect information. In sharp contrast, a carefully designed hypothesis test allows Alice to significantly improve her payoff in $(\Gamma_f^{e,4})$ given a noisy observation of Eve's move, as compared to the no-information case. Since $P_e=2P_a$ in this
example, an increase in Alice's transmit power also implies an
increase in Eve's power, which aids the hypothesis test at Bob and thus Alice has a better
estimate of Eve's move. On the other hand, Eve's hypothesis test does not show the same improvement as $P_a$ increases since the ratio between data and artificial noise power remains virtually the same.

\section{CONCLUSION}\label{sec:concl}
We have formulated the interactions between a multi-antenna
transmitter and a dual-mode eavesdropper/jammer as a novel zero-sum
game with the ergodic MIMO secrecy rate as the payoff function. We derived
conditions under which Nash equilibria exist and the optimal user policies in both pure and mixed
strategies for the strategic version of the game, and we also
investigated subgame-perfect and sequential equilibria in the
extensive forms of the game with and without perfect information.
 Our numerical results showed that a change in a
single parameter set while others remain constant can shift the
equilibrium from a pure to a mixed NE outcome or vice versa.



\newpage

\begin{table}[ht]
\caption{Payoff matrix $\mathbf{R}$ of the strategic form of the MIMO wiretap game.} \label{table:game}
\centering
\begin{tabular}{c}
\begin{game}{2}{2}[\textbf{Alice}][\textbf{Eve}]
& $\textit{Eavesdrop (E)}$ & $\textit{Jam Bob (J)}$ \\
$\textit{Full Power (F)}$ &$R_{FE}$ &$R_{FJ}$ \\
$\textit{Artificial Noise (A)}$ &$R_{AE}$ &$R_{AJ}$
\end{game}\hspace*{\fill}%
\end{tabular}
\end{table}

\newpage

\newpage

\begin{figure}[b!]
\hspace*{\fill}
\begin{egame}(600,380)
\putbranch(300,240)(2,1){200}
\iib{Eve}{$E$}{$J$}
\putbranch(100,140)(1,1){100}
\iib{Alice}[l]{$F$}{$A$}[$R_{FE}$][$R_{AE}$]
\putbranch(500,140)(1,1){100}
\iib{Alice}[r]{$F$}{$A$}[$R_{FJ}$][$R_{AJ}$]
\end{egame}
\hspace*{\fill}
\caption[]{Extensive form game tree with perfect information $\Gamma^{e,2}$ where Eve moves first and Alice moves second.}\label{fig:extensive2}
\end{figure}

\newpage

\newpage

\newpage

\newpage


\begin{thebibliography}{1}
\bibitem{Wyner75}
A. D. Wyner, ``The wire-tap channel," \emph{The Bell Systems Technical Journal}, vol. 54, pp. 1355-1387, 1975.

\bibitem{LeungH78}
S. L. Y. Cheong and M. Hellman, ``The Gaussian wire-tap channel," \emph{IEEE Trans. Inf. Theory}, vol. 24, no.
4, pp. 451-456, July 1978.

\bibitem{Csiszar78}
I. Csisz\'{a}r and J. K\"orner, ``Broadcast channels with confidential messages," \emph{IEEE Trans. Inf. Theory}, vol. 24, no. 3, pp. 339-348, May 1978.
\bibitem{OggierH08}
F. Oggier and B. Hassibi, ``The secrecy capacity of the MIMO wiretap channel," \emph{IEEE Trans. Inf. Theory}, vol. 57, no. 8, pp. 4961-4972, Aug. 2011.
\bibitem{Petropulu11}
J. Li and A. Petropulu, ``On ergodic secrecy capacity for Gaussian MISO wiretap channels," \emph{IEEE Trans. Wireless Commun.}, vol. 10, no. 4, pp. 1176 - 1187, Apr. 2011.
\bibitem{Shitz09}
T. Liu and S. Shamai, ``A note on the secrecy capacity of the multiple-antenna wiretap channel," \emph{IEEE Trans. Inf. Theory}, vol. 55, no. 6, pp. 2547-2553, June 2009.
\bibitem{GoelN08}
S. Goel and R. Negi, ``Guaranteeing secrecy using artificial noise," \emph{IEEE Trans. Wireless Commun}., vol. 7, no. 6, pp.
2180-2189, June 2008.
\bibitem{MISOME}
A. Khisti and G. Wornell, ``Secure transmission with multiple antennas I: The MISOME wiretap channel", \emph{IEEE Trans. Inf. Theory}, vol. 56, no. 7, pp. 3088-3104 , July 2010.
\bibitem{Wornell09}
A. Khisti and G. Wornell, ``Secure transmission with multiple antennas II: The MIMOME wiretap channel", \emph{IEEE Trans. Inf. Theory}, vol. 56, no. 11, pp. 5515-5532, Nov. 2010.

\bibitem{MukherjeeTSP}
A. Mukherjee and A. L. Swindlehurst, ``Robust beamforming for secrecy in MIMO wiretap channels with imperfect CSI," \emph{IEEE Trans. Signal Process.}, vol. 59, no. 1, pp. 351-361, Jan. 2011.
\bibitem{Stark88}
W. E. Stark and R. J. McEliece, ``On the capacity of channels with block memory," \emph{IEEE Trans. Inf. Theory}, vol. 34, no. 3, pp. 322-324, Mar. 1988.
\bibitem{Medard97}
M. M\'edard, ``Capacity of correlated jamming channels," in \emph{Proc. 35th Allerton Conf.}, pp. 1043-1052, 1997.
\bibitem{Diggavi01}
S. N. Diggavi and T. M. Cover, ``The worst additive noise under a covariance constraint," \emph{IEEE Trans. Inf. Theory}, vol. 47, pp. 3072-3081, Nov. 2001.
\bibitem{Basar04}
A. Kashyap, T. Ba\c{s}ar, and R. Srikant, ``Correlated jamming on MIMO Gaussian fading channels," \emph{IEEE Trans. Inf. Theory}, vol. 50, no. 9, pp. 2119-2123, Sep. 2004.
\bibitem{Khandani04}
A. Bayesteh, M. Ansari, and A. K. Khandani, ``Effect of jamming on the capacity of MIMO channels," in \emph{Proc. 42nd Allerton Conf}., pp. 401-410, Oct. 2004.
\bibitem{Ulukus06}
S. Shafiee and S. Ulukus, ``Mutual information games in multi-user channels with correlated jamming," \emph{IEEE Trans. Inf. Theory}, vol. 55, no. 10, pp. 4598-4607 , Oct. 2009.
\bibitem{Giannakis08}
T. Wang and G. B. Giannakis, ``Mutual information jammer-relay games," \emph{IEEE Trans. Inf. Forensics Security}, vol. 3, no. 2, pp. 290-303, June 2008.
\bibitem{Amariucai09}
G. Amariucai and S. Wei, ``Half-duplex active eavesdropping in fast fading channels: A block-Markov Wyner secrecy encoding scheme," submitted to \emph{IEEE Trans. Inf. Theory}, 2010, available: arXiv:1002.1313.
\bibitem{Erkip09}
M. Y\"uksel, X. Liu, and E. Erkip, ``A secure communication game with a relay helping the eavesdropper," \emph{IEEE Trans. Inf. Forensics Security}, vol. 6, no. 3, pg. 818-830, Sep. 2011.
\bibitem{Swindlehurst09}
A. L. Swindlehurst, ``Fixed SINR solutions for the MIMO wiretap channel," in \emph{Proc. IEEE ICASSP}, pp. 2437-2440, 2009.
\bibitem{Mukherjee09}
A. Mukherjee and A. L. Swindlehurst, ``Fixed-rate power allocation strategies for enhanced secrecy in MIMO wiretap channels," in \emph{Proc. IEEE SPAWC}, pp. 344-348, Perugia, June 2009.
\bibitem{Zhou09}
X. Zhou and M. R. McKay, ``Secure transmission with artificial noise over fading channels: Achievable rate and optimal power allocation," \emph{IEEE Trans. Veh. Tech}., vol. 59, no. 8, pp. 3831-3842, Oct. 2010.
\bibitem{Hong_TWC11}
S.-C. Lin, T.-H. Chang, Y.-L. Liang, Y.-W. P. Hong, and C.-Y. Chi, ``On the impact of quantized channel feedback in guaranteeing secrecy with artificial noise -
the noise leakage problem, \emph{IEEE Trans. Wireless Commun}., vol. 10, no. 3, pp. 901-915, Mar. 2011.
\bibitem{ICC10}
A. Mukherjee and A. L. Swindlehurst, ``Equilibrium outcomes of dynamic games in MIMO channels with active eavesdroppers," in \emph{Proc. IEEE ICC}, Cape Town, South Africa, May 2010.
\bibitem{MILCOM10}
A. Mukherjee and A. L. Swindlehurst, ``Optimal strategies for countering dual-threat jamming/eavesdropping-capable adversaries in MIMO channels," in \emph{Proc.  IEEE MILCOM}, San Jose, CA, Nov. 2010.
\bibitem{Ulukus07}
S. Shafiee and S. Ulukus, ``Achievable rates in Gaussian MISO channels with secrecy constraints,''
in \emph{Proc. IEEE ISIT}, 2007.
\bibitem{Petrosjan}
L. A. Petrosjan and N. A. Zenkevich, \emph{Game Theory}. World Scientific, 1996.
\bibitem{Fudenberg}
D. Fudenberg and J. Tirole, \emph{Game Theory}. MIT Press, 1991.
\bibitem{Myerson}
R. Myerson, \emph{Game Theory: Analysis of Conflict}. Harvard University Press, 1997.
\bibitem{Liu09}
Y. Wu, B. Wang, K. J. R. Liu, and T. C. Clancy, ``Repeated open spectrum sharing game with cheat-proof strategies," \emph{IEEE Trans. Wireless Commun.}, vol. 8, no. 4, pp. 1922-1933, Apr. 2009.

\bibitem{Gazor10}
A. Taherpour, M. Nasiri-Kenari, and S. Gazor, ``Multiple
antenna spectrum sensing in cognitive radios," \emph{IEEE
Trans. Wireless Commun.}, vol. 9, pp. 814-823, Feb. 2010.
\bibitem{KayVolII}
S. M. Kay, \emph{Fundamentals of Statistical Signal Processing vol. II- Detection Theory}. Prentice Hall, 1998.
\end{thebibliography}
\end{document}